\documentclass[12pt]{article}
\usepackage{epsf}
\usepackage{amsmath}
\usepackage{graphics}
\usepackage{cite}

\renewcommand{\thefootnote}{\#\arabic{footnote}}
\begin{document}

\newcommand{\gtrsim}{ \mathop{}_{\textstyle \sim}^{\textstyle >} }
\newcommand{\lesssim}{ \mathop{}_{\textstyle \sim}^{\textstyle <} }

\newcommand{\rem}[1]{{\bf #1}}

\renewcommand{\thefootnote}{\fnsymbol{footnote}}
\setcounter{footnote}{0}
\begin{titlepage}

\def\thefootnote{\fnsymbol{footnote}}

\begin{center}

\hfill hep-ph/0402119\\
\hfill February 2004\\

\vskip .5in

{\Large \bf
Candidates for Inflaton in Quiver Gauge Theory }

\vskip .45in

{\large
Paul H. Frampton and Tomo Takahashi
}

\vskip .45in

{\em
Department of Physics and Astronomy,\\
University of North Carolina, Chapel Hill, NC 27599-3255, USA
}

\end{center}

\vskip .4in

\begin{abstract}
The inflaton necessary to implement the mechanism of inflationary
cosmology has natural candidates in quiver gauge theory.  We discuss
the dimensionless coefficients of quartic couplings and enumerate
scalars which are singlet under the low-energy gauge group.  The
inflaton mass is generally predicted to be in the TeV region, close to
4 TeV for one specific unified model.  A quartic inflaton potential,
and a mutated hybrid inflation, are discussed. They can give adequate
inflation and appropriate fluctuations but different spectral indices.
\end{abstract}
\end{titlepage}

\renewcommand{\thepage}{\arabic{page}}
\setcounter{page}{1}
\renewcommand{\thefootnote}{\#\arabic{footnote}}

\section{Introduction}

The unadorned Big Bang model has well-known shortcomings, particularly
the extreme fine-tuning of initial conditions necessary to accommodate
the flatness and horizon issues.  Inflation \cite{Guth:1980zm} is the
leading candidate to augment the Big Bang and it is hardly necessary
to justify here its successes, not only in solving the flatness and
horizon problems but also in providing a plausible origin of the
primordial fluctuations which lead to structure formation.

Inflation, however, requires the presence of one or more scalar fields
whose potential underlies the mechanism for early era of 
super-rapid cosmic expansion.  In the minimal
standard model (SM) of particle phenomenology the only scalar is the
Higgs boson which cannot be the inflaton for the obvious reason that
it would not respect the gauge invariance of the SM in the early
universe.  Usually, therefore, one or more additional scalar fields
are postulated to set up the inflaton potential in a way unconnected
to the SM. Some partially successful efforts
\footnote{For a comprehensive review, see \cite{LR}}
have been made to identify the inflaton
within supersymmetry theory {\it e.g.} \cite{DR} and
in superstring theory {\it e.g.} \cite{GLM}.

In this article we study how quiver gauge theories which have been
suggested as attractive extensions of the SM up to TeV energies may at
the same time provide natural candidates for the scalars necessary for
the inflationary scenario in the early universe.

It will turn out that there is a plethora of possible scalars to play
the inflaton role and the quartic potential describing them has a lot
of flexibility in its coefficients.  Here we merely make a preliminary
exploration of the types of inflaton potential which are suggested by
the quiver gauge theory.

\bigskip
\bigskip

\section{Quiver Gauge Theories}

It has been argued that the appropriate way to extend
the standard model to higher energies in the TeV
regime, as an alternative to either supersymmetry
or technicolor, is to embed it in a nonsupersymmetric
quiver gauge theory. It is conjectured that
such a theory achieves naturalness in the
sense of absence
of quadratic divergences\cite{JJ,PHF0}

Such theories have been described in detail elsewhere
so suffice it to characterize two types of abelian quiver ${\cal N}
=0$ theories, based on $Z_7$ and $Z_{12}$ respectively.

The theories are specified by the embedding of the abelian
finite group $\Gamma = Z_p$ in the
$SU(4)$ global symmetry of the
ancestral ${\cal N} = 4$ U(N) super Yang-Mills (SYM) model.

The embedding is defined by the {\bf 4} of SU(4)
but for the scalar sector it is adequate to cite the corresponding {\bf 3}
in the {\bf 6 = 3 + $3^*$}. Thus for the three viable
$Z_7$ theories of \cite{PHF1} we have respectively:

\noindent $a_i = (2, 3, 3)$ (Model 7B),
$a_i = (1, 1, 3)$ (Model 7D) and $a_i = (1, 2, 2)$ (Model 7E).

\bigskip

The corresponding labeling of the heptagonal quiver
nodes are:

\noindent (7B) - C - W - H - W - H - H - H -

\noindent (7D) - C - W - H - H - H - W - H -

\noindent (7E) - W - H - H - H - W - W - H -

\noindent where the C, W, H refer to the three gauge
groups in $SU(3)_C \times SU(3)_W \times SU(3)_H$ trinification.
The two ends of the quivers are in each case identified.

\bigskip

In the $Z_{12}$ model of \cite{PHF2}
the embedding is {\bf 3} = (3, 4, 5) and the
node identification on the dodecahedral quiver is

- C - W - H - C - $W^4$ - $H^4$ -
 
\bigskip

In this $Z_{12}$ model the conformal scale is about 4 TeV (more
precisely 3.8 TeV). In the $Z_7$ models it also must be no too far
above the weak scale to allow the correct value of 
the electroweak mixing angle sin$^2 \theta$ to
survive, so here again it is at the TeV scale.

\bigskip
\bigskip

\section{Standard Model Singlets}

The inflaton fields must necessarily be singlets under the SM gauge
group to avoid unwanted symmetry breaking above the weak scale if the
inflation involves scalar values which are at higher energy.  The
bifundamental scalars of the $Z_p$ quiver model all contain eighteen
real fields and transform under a pair of the gauge groups in the
trinification $SU(3)_C \times SU(3)_W \times SU(3)_H$.

\bigskip

\noindent We analyze each of the six possible pairs from C, W, H in turn:

\bigskip

\noindent {\bf \underline{(C,C)}}

\bigskip

\noindent 
This bifundamental transforms under color $SU(3)_C$
as 2({\bf 1 + 8}) and carries no $SU(2) \times U(1)$
quantum numbers.

\bigskip

\noindent {\it $(C, C)$ contains two 321-singlets.}

\bigskip

\noindent {\bf \underline{(C,W) + (W,C)}}

\bigskip

\noindent Using the weak hypercharge formula\cite{FRT}

\begin{equation}
Y = \frac{2}{\sqrt{3}} \left( T_{8W} - T_{8H} \right)
\label{Y}
\end{equation}

\noindent 
with $(2/\sqrt{3}) T_8 = (1/3)$ diag (1, 1, -2)
we see that all the components have non-zero $Y$.

\bigskip

\noindent {\it $(C, W) + (W, C)$  contains no 321-singlet.}

\bigskip

\noindent {\bf \underline{(W,W)}}

\bigskip

\noindent 
This corresponds to 2({\bf 8 + 1}) under $SU(2)_L$,
singlet under $SU(3)_C$ and $U(1)_Y$.

\bigskip
\bigskip

\noindent {\it $(W, W)$ contains two 321-singlets.}

\bigskip

\noindent {\bf \underline{(C,H)+(H,C)}}

\bigskip

\noindent 
Similarly to  $(C, W) + (W, C)$,
all components have non-vanishing $Y$, and triplet color.

\bigskip

\noindent {\it (C, H) + (H, C) contains no 321-singlet.}

\bigskip

\noindent {\bf \underline{(W,H)+(H,W)}}

\bigskip

\noindent 
There are several ways to see that this has two 321-singlets.
Think of the {\bf 27} of $E(6)$ which contains
{\bf $27 = 10 + \bar{5} + (5 + \bar{5}) + 1 + 1$}
under $SU(5)$. One can confirm this in \cite{Slansky}.

\bigskip

\noindent {\it $(W, H) + (H, W)$ contains two 321-singlets.}

\bigskip

\noindent {\bf \underline{(H,H)}}

\bigskip

\noindent 
This is 2({\bf 8 + 1}) and has no color
or weak isospin. The {\bf 1} has vanishing $Y$. 

\bigskip

\noindent {\it $(H, H)$ contains two 321-singlets.}

\bigskip

\section{Counting Candidate for Inflatons}

Let us first consider a general model which contains the SM. 
The node identification on the
quiver is some pattern of C, W, and H
and the scalar embedding is $a_i = (a_1, a_2, a_3)$. If we consider
the scalar bifundamentals emanating
from the node $p$ there are six of them ($i = 1, 2, 3$):

\begin{equation}
 \left[ (3_p, 3^*_{p+a_i}) ,  (3_{p-a_i}, 3^*_p) \right]
\label{pnode}
\end{equation}
and we can label these six fields as $\Phi^{(p)}_i, \Phi^{(p)'}_i$
respectively. We now define linear combinations:

\begin{equation}
\Psi^{(p)}_k 
= \sum_i \left[ 
\lambda^{(p)}_{ki} \Phi_i + \lambda^{(p)'}_{ki} \Phi^{'}_i 
\right]
\label{Psi}
\end{equation}

From\cite{BFLP} the relevant potential is summing $p$ from 1 to 7:

\begin{equation}
V = 2g_0^2 \sum_{p=1}^{7}
\left[ \left(\sum_{i=3}^{3} | (3_p, 3^*_{p+a_i}) |^2 \right)^2
-\sum_{i,j = 1}^3 | (3_{p - a_i}, 3^*_{p}) |^2
| (3_{p}, 3^*_{3 + a_j} ) |^2 \right]
\label{V}
\end{equation}
together with other terms not relevant to this discussion.

By studying the linear combinations in Eq.(\ref{Psi}) and asking
that in Eq.(\ref{V}) the coefficient of $|\Psi_k|^4$
be arbitrarily adjustable, it can be shown that for each fixed node
number $p$ there are exactly five independent $\Psi_k$
which correspond to the desirable directions.

However, a second requirement for a candidate inflaton is that
it be a 321-singlet. This uses the analysis
of the previous Section and is model dependent.

Let us first examine the $Z_7$ models in \cite{PHF1}.
For the case (7B),  all node except the one C node
generate five candidate directions so the total number is
$6 \times 5 = {\bf 30}$. The same number of such 
directions is available for the others $Z_7$
cases (7D) and (7E).

\bigskip

In the $Z_{12}$ model, the two C nodes do not generate
appropriate directions. By looking at 321-singlets, there is
also a reduction by one direction for
each of one W and one H node which bisect the two
C nodes on the opposite side of the quiver.
The total number of candidate inflatons for $Z_{12}$
is therefore $10 \times 5 - 2 = {\bf 48}$.

For the scalars so enumerated, linear combinations can
give arbitrary coefficients in the resultant quartic
inflaton potential.

\bigskip
\bigskip

\section{Inflaton Potential}

Now we consider an inflaton potential arising from the quiver gauge
theory. Since there are many quartic terms in the scalar potential,
considerations in the previous sections lead initially to the
following simplest possibility for the inflaton potential:

\begin{equation}
V(\Phi) = 
V_0 + \frac{1}{4} \lambda \Phi^4
\label{potential}
\end{equation}
where $V_0 = M_{\rm CSB}^4$. $M_{\rm CSB}$ is the scale at which the
conformal symmetry breaks down.  Since the energy scale at which
conformality is broken lies between the unification scale $\Lambda
\simeq 4 $ TeV and the (reduced) Planck scale $M_{\rm Planck} \simeq
2.4 \times 10^{18}$ GeV, we assume $ \Lambda < M_{\rm CSB} < M_{\rm
Planck}$ in the following arguments. Furthermore, the scalar fields
which are inflaton candidates must not develop vacuum values $<\Phi>$
which are greater than $\Lambda$ during inflation to avoid breaking
the $SU(N)^p$ gauge symmetry. Thus such values are very much smaller
than the inflation scale $<\Phi> \ll M_{CSB}$.

Having these consideration in mind, now we look into the inflation
with the potential given by Eq.\ (\ref{potential}).  In the slow-roll
approximation, the number of e-foldings $N$ during inflation is given
by \cite{LR}:

\begin{equation}
N \simeq \frac{1}{M_{\rm Planck}^2} \int_{\Phi_{\rm end}}^{\Phi}
d\Phi \frac{V}{V^{'}}
\label{N}
\end{equation}
where  a prime denotes a derivative with respect to $\Phi$.
For the potential Eq.\ (\ref{potential}) $N$ becomes

\begin{equation}
N = \frac{V_0}{2\lambda M_{\rm Planck}^2} \left[ \frac{1}{\Phi_{\rm end}^2} -
\frac{1}{\Phi^2} \right] 
\simeq 
\frac{1}{2\lambda}  \left(\frac{M_{\rm CSB}}{\Phi_{\rm end}} \right)^2
\left(\frac{M_{\rm CSB}}{M_{\rm Planck}} \right)^2
\label{N2}
\end{equation}
where we assumed $\Phi_{\rm end} \ll \Phi$.

If we put a coupling $\lambda \sim 1$ then for scalar values below TeV
scale and the conformality symmetry breaking scale $M_{\rm CSB}
\gtrsim \Lambda$, we can have the sufficient amount of inflation which
is needed to solve the horizon and flatness problems.  Now we consider
the density perturbation. One of the major virtues of the inflation is
that it can generate the density fluctuation which is the origin of
cosmic microwave background anisotropy and large scale
structure\footnote{
Other mechanisms to generate the density perturbation have been
proposed such as the curvaton mechanism \cite{curvaton} and the
inhomogeneous reheating scenario \cite{inhomdecay}.  However we do not
consider such scenarios here.}.
The normalization\footnote{
Although we do not need the precise value of $\delta_H$ for the
purpose of this paper, we give the value from WMAP
\cite{Spergel:2003cb}.  The observation of WMAP gives $
(25/4)\delta_H^2 = 2.95 \times 10^{-9} A$ with $A=0.9\pm 0.1$ at
$k=0.05 {\rm Mpc}^{-1}$ for a power-law $\Lambda$CDM model. }
of the spectrum at the COBE scale is given by $\delta_H \sim 2 \times
10^{-5}$ where $\delta_H$ can be written as

\begin{equation}
\delta_H = \frac{2}{5} P_R = \frac{1}{\sqrt{75\pi^2}} 
\frac{V^{3/2}}{M_{\rm Planck}^3 V'}
\label{deltaH}
\end{equation}
With the potential Eq.\ (\ref{potential}), 
\begin{equation}
\frac{V^{3/2}}{M_{\rm Planck}^3 V^{'}} 
\simeq
\frac{M_{\rm CSB}^6}{\lambda \Phi_{\rm COBE}^3 M_{\rm Planck}^3}  
\label{COBE}
\end{equation}
which should be $5 \times 10^{-4}$.

Since $\Phi_{\rm end} \leq \Phi_{\rm COBE}$ in this model,
from Eq.\ (\ref{N2}), 
\begin{equation}
\frac{V^{3/2}}{M_{\rm Planck}^3 V^{'}} 
\geq
\sqrt{8\lambda N^3}
\label{COBE2}
\end{equation}
For $\lambda \sim 1$, $\Phi\leq 1$ TeV, and using\footnote{
Actually, the number $N$ depends on the thermal history of the
universe as \cite{LR}
\begin{equation}
N = 62 - \ln \left(\frac{k}{a_0 H_0} \right)-
\ln \left(\frac{10^{16} {\rm GeV}}{V_k^{1/4}} \right)+ 
\ln \left(\frac{V_k^{1/4}}{V_{\rm end}^{1/4}} \right)
-\frac{1}{3} \ln \left( \frac{V_{\rm end}^{1/4}}{\rho_{\rm reh}^{1/4}} \right), 
\end{equation}
where the subscripts ``$k$,'' ``end'' and ``reh'' indicate the value of 
quantities when $k=aH$, the inflation ends and after reheating respectively.
For the purpose of this article, we can take $N=50$.
}
$N \simeq 50$, the fluctuation will be too large as seen from Eq.\
(\ref{COBE2}).  However we can make use of defining linear
combinations of fields which appear in the theory, and can use
different values (e.g. $\lambda \sim 10^{-13}, M_{\rm CSB} \sim 10^4$
TeV). With these parameter values, we can have the right amount of
fluctuation. In this potential, the slow-roll parameters are
\begin{eqnarray}
\epsilon &=& \frac{1}{2} \lambda^2
\left(\frac{M_{\rm Planck}}{M_{\rm CSB}} \right)^2 
\left(\frac{\Phi}{M_{\rm CSB}} \right)^6,  \\
\eta     &=& 3 \lambda 
\left(\frac{M_{\rm Planck}}{M_{\rm CSB}} \right)^2 
\left(\frac{\Phi}{M_{\rm CSB}} \right)^2. 
\end{eqnarray}
Since $\epsilon$ can be neglected compared to $\eta$, we have 
\begin{equation}
n-1 \simeq 2\eta = 6 \lambda 
\left(\frac{M_{\rm Planck}}{M_{\rm CSB}} \right)^2 
\left(\frac{\Phi}{M_{\rm CSB}} \right)^2
\end{equation}
which is a slightly blue-tilted spectrum. For example, if we choose
the parameters as $\lambda \sim 10^{-13}, M_{\rm CSB} \sim 10^4$ TeV and 
$\Phi \sim 100$ GeV, we get $n-1 \sim 0.03$.

We can have other fields in the general scalar potential.  Thus, next
we consider the possibility of an inflaton potential which involves
two scalar fields.

\bigskip
\bigskip

{\it Mutated Hybrid Inflation}

\bigskip
\bigskip

From the consideration in the previous section, we can have a more
general potential involving two scalars which can be written as

\begin{equation}
V(\Phi,\Psi) = V_0 - \frac{1}{2}m^2\Psi^2 + \frac{1}{3}\lambda' \Psi^3 \Phi
\label{V2}
\end{equation}
An inflation model with this kind of potential is known as ``mutated
hybrid inflation'' \cite{LR}.  n the potential above, for some fixed
$\Phi$, there exists a minimum for $\Psi_c = -m^2/\lambda' \Phi$.
Substituting $\Psi_c$ into the potential, $V$ can be written as

\begin{equation}
V = V_0 (1 - \mu \Phi^{-2})
\label{V3}
\end{equation}
where $\mu = m^6/(6\lambda'^2 V_0)$.  Now we consider the number of
e-foldings and density perturbation from this potential. Using Eq.\
(\ref{N}), The number of e-foldings is given by:

\begin{equation}
N \simeq \frac{\Phi^4}{8 M_{\rm Planck}^2 \mu}.
\label{N4}
\end{equation}
With this equation, the value of $\Phi$ during inflation can be
written with $N$.  The density perturbation generated by this
potential is, Using Eqs.\ (\ref{deltaH}) and (\ref{N4}),

\begin{equation}
\frac{V_0^{3/2}}{M_{\rm Planck}^3 V'} 
= \frac{\sqrt{V_0}\Phi^3}{2\mu M_{\rm Planck}^3} 
= \left(\frac{M_{\rm CSB}}{M_{\rm Planck}} \right)^2 \mu^{-1/4}
\left(32 M_{\rm Planck}^2 N^3 \right)^{1/4}
\sim 5 \times 10^{-4}.
\label{fluct}
\end{equation}
From Eq.\ (\ref{N4}), assuming $\Phi \sim 1$ TeV and using $N \simeq
50$, we can fix the value of $\mu$.  Putting this value into Eq.\
(\ref{fluct}), we find $M_{\rm CSB} \simeq 10^5$ TeV which is
reasonable value for $M_{\rm CSB}$. In other words, choosing the value
of $\mu$ appropriately, this model gives the sufficient e-folding
number as well as right amount of the density perturbation.

To see the spectral index of the scalar perturbation, first we write 
down the slow-roll parameters
\begin{eqnarray}
\epsilon &=& \frac{2 M_{\rm Planck}^2 \mu^2}{\Phi^6} 
= \frac{1}{32N^2}\left( \frac{\Phi}{M_{\rm Planck}} \right)^2, \\
\eta     &=& -\frac{6 M_{\rm Planck}^2 \mu}{\Phi^4} = -\frac{3}{4N}.
\end{eqnarray}
The spectral index is 
\begin{equation}
n-1 = 2\eta -6\epsilon \simeq -\frac{3}{2N}
\end{equation}
where we neglect $\epsilon$ since it is negligible for $\Phi \lesssim
\Lambda$.  Hence this model predicts a slightly red-tilted spectrum
which is consistent with current observations.  Furthermore, because
of the low inflation scale in this model, the contribution of
gravitational waves is negligible, which is also consistent with
observations.

\bigskip
\bigskip

\section{Discussion}

Here we have made an attempt to connect particle phenomenology	
to the inflationary era in the early universe.

In CFT there is no shortage of candidates for the scalar fields to be
involved in the inflaton potential. By taking linear combinations of
scalar fields which are ubiquitous in the theory thereof we can arrive
at a quartic potential with quite general coefficients with arbitrary
sign.

Taking the simplest single-inflaton potential\footnote{
 We have provisionally assumed that 1-loop contributions to $V$ are
negligible although this assumption depends on the details of
conformal symmetry breaking as well as renormalization properties of
quiver gauge theories which are currently under study.}
we arrived at sufficient inflation by assuming the conformal symmetry
breaking scale is much higher than the TeV scale and that the scalar
values are around the TeV scale.  In this case, primordial
fluctuation with a slightly blue-tilted spectrum is predicted.

By taking a more general mutated hybrid potential involving two scalar
fields, not only an adequate amount of inflation was achieved but also
the correct size for the primordial fluctuations with a slightly
red-tilted spectrum. Again this assumed that the conformal symmetry
breaking scale is at about $10^5$ TeV while the scalar values remain
at or below a TeV scale.

Thus we conclude that quiver gauge theory does contain natural
candidate(s) for inflaton(s).

\section*{Acknowledgment}
This work provides an answer to a question raised by Professor David
H.~Lyth after a conference talk by one of us (PHF) in Summer 2003.
This work was supported in part by the US Department of Energy under
Grant No. DE-FG02-97ER-41036.

\end{document}